\documentclass[5p,times,twocolumn,sort&compress]{elsarticle}

\usepackage{lineno,hyperref}
\modulolinenumbers[5]

\usepackage{pifont}
\usepackage{natbib}
\usepackage{geometry}
\usepackage{graphicx}
\usepackage{txfonts}
\usepackage{hyperref}
\usepackage{bm}
\usepackage{color}

\newcommand{\dd}{\mathrm{d}}
\newcommand{\diff}[2]{\frac{\dd #1}{\dd #2}}

\journal{Physics Letters B}

\begin{document}

\title{Extrasolar planets as a probe of modified gravity}

\author[if,on,ita]{Marcelo Vargas dos Santos}
\ead{mvsantos@protonmail.com}

\author[ita]{David F. Mota}
\ead{d.f.mota@astro.uio.no}

\address[if]{Instituto de F\'isica, Universidade Federal do Rio de Janeiro, 21941-972, Rio de Janeiro, RJ, Brazil}
\address[on]{Departamento de Astronomia, Observat\'orio Nacional, 20921-400, Rio de Janeiro – RJ, Brasil}
\address[ita]{Institute of Theoretical Astrophysics, University of Oslo, Postboks 1029, 0315 Oslo, Norway}

\begin{abstract}
We propose a new method to test modified gravity theories, taking advantage of the available data on extrasolar planets. We computed the deviations from the Kepler third law and use that to constrain gravity theories beyond General Relativity. We investigate gravity models which incorporate three screening mechanisms: the Chameleon, the Symmetron and the Vainshtein. We find that data from exoplanets orbits are very sensitive to the screening mechanisms putting  strong constraints in the parameter space for the Chameleon models and the Symmetron, complementary and competitive to other methods, like interferometers and solar system. With the constraints on Vainshtein we are able to work beyond the hypothesis that the crossover scale is of the same order of magnitude than the Hubble radius $r_c \sim H_0^{-1}$, which makes the screening work automatically, testing how strong this hypothesis is and the viability of other scales.
\end{abstract}

\begin{keyword}
	Gravity \sep Modified Gravity \sep General Relativity \sep Fifth Force \sep Extrasolar Planets \sep Exoplanets

\end{keyword}

\maketitle

\section{Introduction} Gravity theories beyond General Relativity (GR) are a possible theoretical framework to explain several cosmological problems \cite{Clifton:2011jh}. In particular, the intriguing present day's cosmic accelerated expansion \cite{Riess:1998cb,Perlmutter:1998np}. Theoretical models which predict an extension to General Relativity must, however, comply with strong requirements: first of all the model must have similar predictions to those of the benchmark model $\Lambda$CDM at cosmological scales: observational data from both the background evolution and the linear large scale structure formation regime is fully consistent with $\Lambda$CDM \cite{Ade:2015xua}. Another condition is that the modifications to General Relativity must be suppressed, by physical mechanisms, in the regimes which are well tested, e.g. solar system scales. This requirement is assured via the so called screening mechanisms \cite{Joyce:2014kja}.

Modified gravity models with screening mechanism have been extensively studied in the literature: either focusing on the background cosmology \cite{Hinterbichler:2011ca,Khoury:2003rn}, large and linear cosmological scales \cite{mota1, mota2, mota3, mota4} or on astrophysical scales in the nonlinear regime \cite{mota5}, and finally at the small solar system scales using local gravity tests \cite{Bertotti:2003rm,Li:2014hya}.

Any weakness in the screening mechanisms should result in appreciable deviations in what we predict from the General Relativity or, in weak field regime, from the Newtonian gravity. This kind of deviations have been used to test modified gravity inside the solar system with, for instance, spectral deviation data from the Cassini space mission, which ensures that the gravitational potential at the Sun surface must deviates less than $10^{-5}$ from the value predicted by the Newtonian gravity \cite{Bertotti:2003rm}. Another interesting work investigates how much the gravity may deviates from the Newtonian gravity using the well measured solar system bodies orbits \cite{Li:2014hya}.

An important feature in the screening mechanisms is the dependence on the physical properties of the environment as the density field, for instance. Thus we expect that the information from different planetary systems should give more statistical significance once the screening works in different ways for each one. On the other side any significantly deviation should already be measured in Solar System, therefore this deviation must be small even in other planetary system.

In this work we investigate the possibility of using exoplanet data to test and constrain modified gravity models. This is not the first attempt to constrain modified gravity with exo\-pla\-nets \cite{Iorio:2009ek}, but in there the authors just compared the theoretical prediction with only one measurement, the transiting e\-xo\-pla\-net HD209458b ``Osiris''. This lacks statistical rigor, and in this work, we use more than one hundred data points and propose a statistical method to make a thorough treatment and study of these systems. The data are obtained from \url{exoplanets.org} portal \cite{2014PASP..126..827H}.

\section{The method} For any gravitational theory the planetary motion is described by the dynamics of a particle under the influence of a central force, i.e. the spatial dependence of the force is only on the distance of the planet to the force center, inside the host star. The relation between the revolution period, $T$, of a planet in a circular orbit, of radius $r$, and the absolute value of the gravitational force, $F(r)$, is given by \citep{PoissonWill201407}\begin{equation}
	T^2 = \frac{4 \pi^2 r}{F(r)} \left( \frac{1}{M_\star} + \frac{1}{M_\mathrm{p}}\right)^{-1},
\end{equation}where $M_\mathrm{p}$ and $M_\star$ are the masses of the planet and the star respectively. For a modified gravity the total force is the sum between the Newtonian force and a fifth force, $F(r) = F_N(r) + F_{5th}(r)$. If the fifth force is null this relation reduces to the third Kepler law\begin{equation}
	T_K^2 = \frac{4\pi^2 r^3}{G(M_\mathrm{p}+M_\star)}.
\end{equation}where $G = 6.67384\times 10^{-8}\mathrm{cm^3g^{-1}s^{-2}}$ is the Newton gravitational constant. So the deviation of the square period from the third Kepler law is\begin{equation}
	\left( \frac{T_K}{T} \right)^2 - 1 = \frac{F_{5th}}{F_N} = \varepsilon.
\end{equation}Therefore, we can use the measured values of $\varepsilon$ to constrain modified gravity using a $\chi^2$ given by the sum between the weighted residuals of all the $N$ measurements\begin{equation}
	\chi^2(\boldsymbol{\theta}) = \sum_{i=1}^N \left(\frac{ \varepsilon_{th}(\phi(\mathbf{x}_i,\boldsymbol{\theta}))-\varepsilon_{obs,i}}{\sigma_i}\right)^2,\label{chi2}
\end{equation}

where $\varepsilon_{th}$ is the theoretical prediction for $\varepsilon$ (the ratio between the fifth and Newtonian forces predicted by theory), $\varepsilon_{obs}$ is the observed value of $\varepsilon$ (the deviation of the square period from the third Kepler law computed from the data), and $\sigma_i$ is the standard deviation (computed from the data by error propagation, the theoretical error can be neglected because it is proportional to $\varepsilon^2$.). $\boldsymbol{\theta}$ is a vector of model parameters and $\mathbf{x}$ is a vector of the physical properties of the star-planet system, which are: $r$ - the planet orbit radius, $R_\star$ - the star radius, $\rho_\star$ - the star density, $\Phi_S$ - the surface Newtonian potential. The field is also a function of the galaxy density, $\rho_g = 10^{-24}g/cm^3$.

To compute the credible regions with 95\% of confidence level (\textrm{C.L.}), we find the values of $\chi^2$ which delimit the bounds ($\chi_b^2$), i.e. the value which gives
\begin{equation}
	P(\boldsymbol{\theta}) = \frac{1}{(2\pi)^{N/2}\prod_i^N\sigma^i} \int_{\Delta\chi^2<\Delta\chi_b^2} \mathrm{e}^{-\Delta\chi^2(\boldsymbol{\theta})/2}\dd\boldsymbol{\theta},\label{prob}
\end{equation}
equals to $0.95$. Where $\Delta\chi^2 = \chi^2-\chi_\mathrm{min}^2$, and $\chi_\mathrm{min}^2$ is the minimum value of $\chi^2$. Fig. \ref{data_dist} (top) shows a comparison between the residuals distribution with a normal distribution of the same mean and standard deviation, which suggests that this is a good approximation for the data distribution. Therefore assuming this distribution to solve \ref{prob} we find $\Delta\chi_b^2\simeq 5.99$ and $\Delta\chi_b^2\simeq 8.08$ for models with 2 and 3 free parameters, respectively \citep{PhilGregory2005}.

The values of $\chi_\mathrm{min}^2$ can me found minimizing the function \ref{chi2}, but this is not a single point for the tested models, there is a degeneracy between the parameters. To avoid this problem we assume $\chi_\mathrm{min}^2$ equal to the value of $\chi^2$ for which the fifth force is null, i.e. $\varepsilon_{th}\equiv 0$, this assumption does not change the results. For a constant deviation, for example, we find $\varepsilon = (0.0_{-6.0}^{+6.0})\times 10^{-3}$ with this assumption and $\varepsilon = (-0.1_{-5.9}^{+6.1})\times 10^{-3}$ without it. A shift less than 1\% of the confidence interval, and it is reduced in the cases with screening.

\section{The data} Our observational data comes from the website \url{exoplanets.org} \citep{2014PASP..126..827H}, which has a compilation of all observed exoplanetary systems: there are 2926 planets with well defined orbits. From those we pick for our analysis 177 planets with circular orbits and with measurements of all properties listed above. These systems are typically composed by a star, similar to the Sun, with a mass that varies from $0.5M_\odot$ to $1.5M_\odot$, and planets like the jovian planets, with masses between $3.2M_\oplus$ and $600M_\oplus$ and typically close to the host star, with orbit radius smaller than $0.5\mathrm{AU}$. This corresponds to short periods, less than a few months.

\begin{figure}[tbp]
\includegraphics[width=\columnwidth]{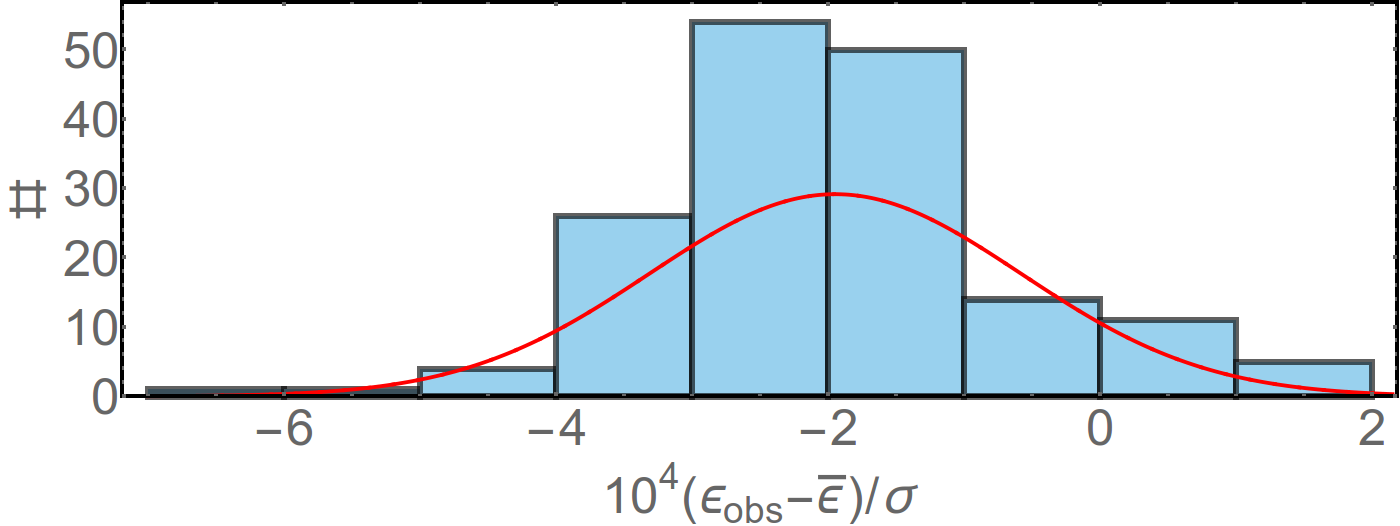}
\includegraphics[width=\columnwidth]{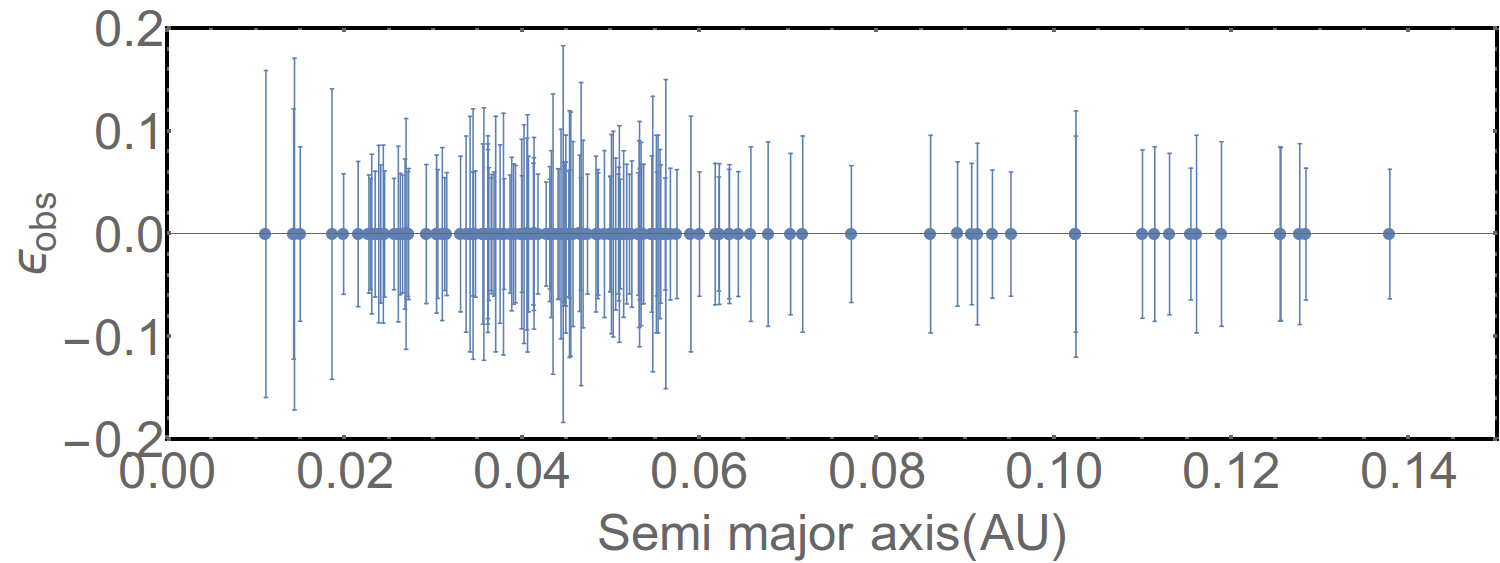}
\caption{Top: The residual distribution of $\varepsilon_{obs}$ compared with a normal distribution with the same mean and standard deviation. Bottom: $\varepsilon_{obs}$ with error bars in function of the semi major axis. However the values are very close to $0$ ($\sim 10^{-5}$) the errors are much larger ($\sim 10^{-1}$), which permits a the existence of a fifth force.}\label{data_dist}
\end{figure}

All the properties are measured by gravity-independent methods, except the orbit radius which use the third Kepler law \citep{Howard:2009pu}. However this looks like a circularity it does not affect considerably our analysis, any appreciable deviation in orbit radius such would already have been measured in the Solar System. The Cassini mission \citep{Bertotti:2003rm} measured, by light time delay, the possible deviation from Newtonian gravitational constant in Solar System and the obtained value is very short
\begin{equation}
	\frac{\Delta G}{G} = \frac{\gamma-1}{2} \lesssim 10^{-5}, ~~~~(0.68\%~\mathrm{C.L.}).
\end{equation}
However the photons may be coupled differently to the field, we expect a correction of the same order in the orbit radius, which is much smaller than $\sigma_i$. Fig. \ref{data_dist} (bottom) shows that the deviations from the third Kepler law ($\varepsilon_{obs}\sim 10^{-5}$) are very smaller than the errors ($\sigma \sim 10^{-1}$), suppressing any possible bias. In summary the possibility to measure deviations from Newtonian gravity is not in the measurements {\it per se} but with their errors. The relativistic corrections, which are less than $10^{-8}$, are not appreciable, in solar system, for example, the most affected body is Mercury, which the orbit radius is well determined by the Newtonian gravity, the effect is seen only in the precession of the perihelion. The orbits of all the exoplanets used in this work are circular ones, therefore there is no perihelion to be precessed.

\section{Modified gravity models}

The models described above were developed in order to explain the cosmic accelerated expansion, so for that reason most of that was designed in a such way that the screening may be automatically satisfied if the background evolution reproduces the same as the benchmark model $\Lambda$CDM. In other words the effective evolution of the Hubble parameter is
\begin{equation}
	H_{eff}^2(z) \simeq H_0^2[\Omega_{m0}(1+z)^3 + 1 - \Omega_{m0}],
\end{equation}
where $H_0$ and $\Omega_{m0}$ are the Hubble constant and the present value of the matter density parameter, respectively. Ones of the most confident measures are that made by the Plack mission \cite{Ade:2015xua}, which are $H_0 = (67.6 \pm 1.0)\rm{km~s^{-1}Mpc^{-1}}$ and $\Omega_{m0} = 0.310 \pm 0.008$. We use that values in our analysis to reduce the number of free parameters in the chameleon models.

In planetary scales we are interested in the correction to the two-body potential energy to find the correction to the third Kepler law. For a complete description of these spherical solutions see \citep{Hinterbichler:2010es,Khoury:2013tda}. Here we only present a brief description of the theories and show the theoretical prediction for the ratio between the fifth force and the Newtonian force, which is needed for our method as mentioned above.

\begin{figure}[bp]
\includegraphics[width=\columnwidth]{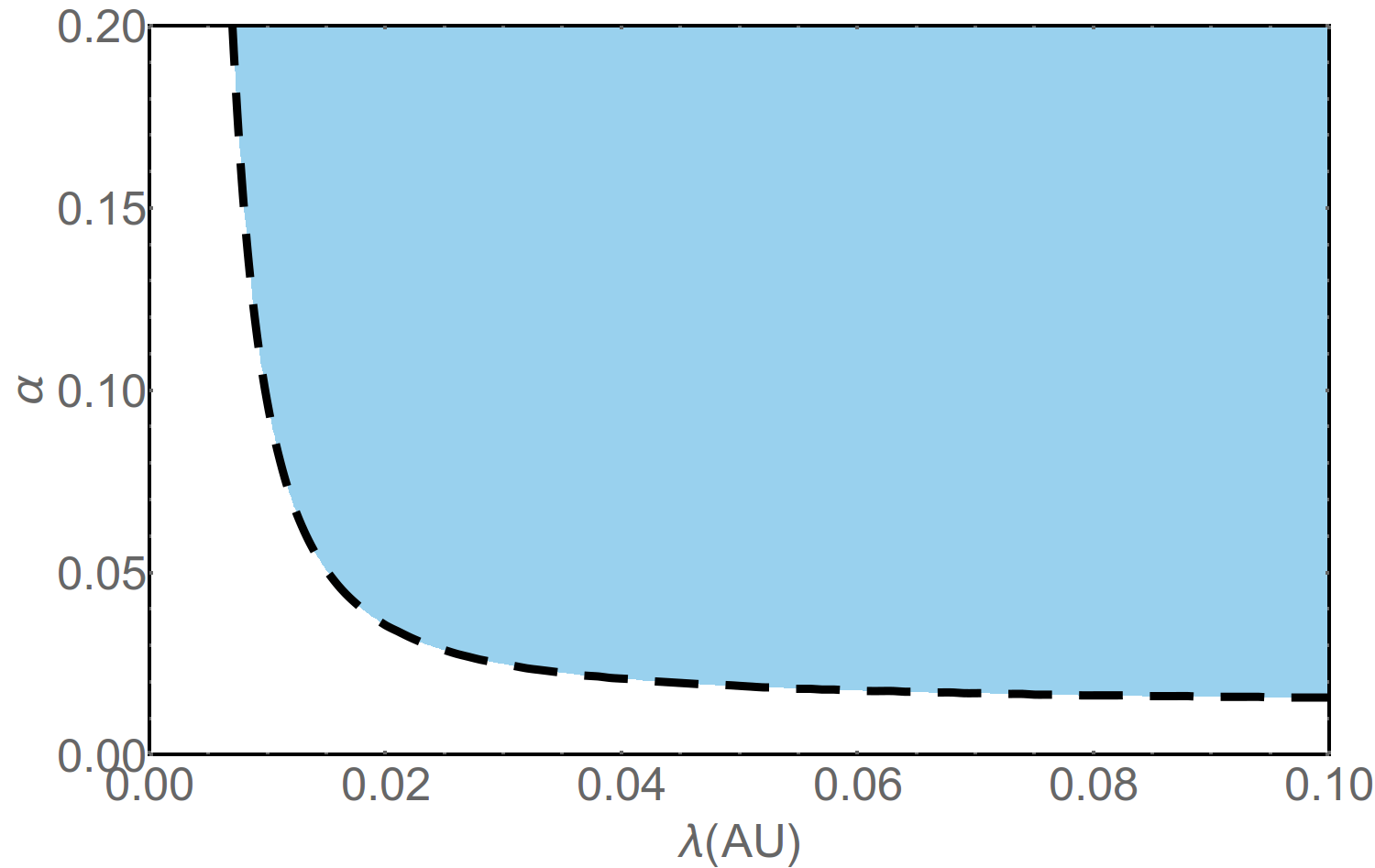}
\caption{Constraints of the unscreened Yukawa potential in ($\lambda$, $\alpha = 2\beta^2$) parameter space, the blue region is ruled out by 95\% of confidence level (95\% \textrm{C.L.}). The GR is recovered when $\beta \to 0$ or $\lambda \to 0$.}\label{yukawa}
\end{figure}
\begin{figure}[tbp]
\includegraphics[width=.9\columnwidth]{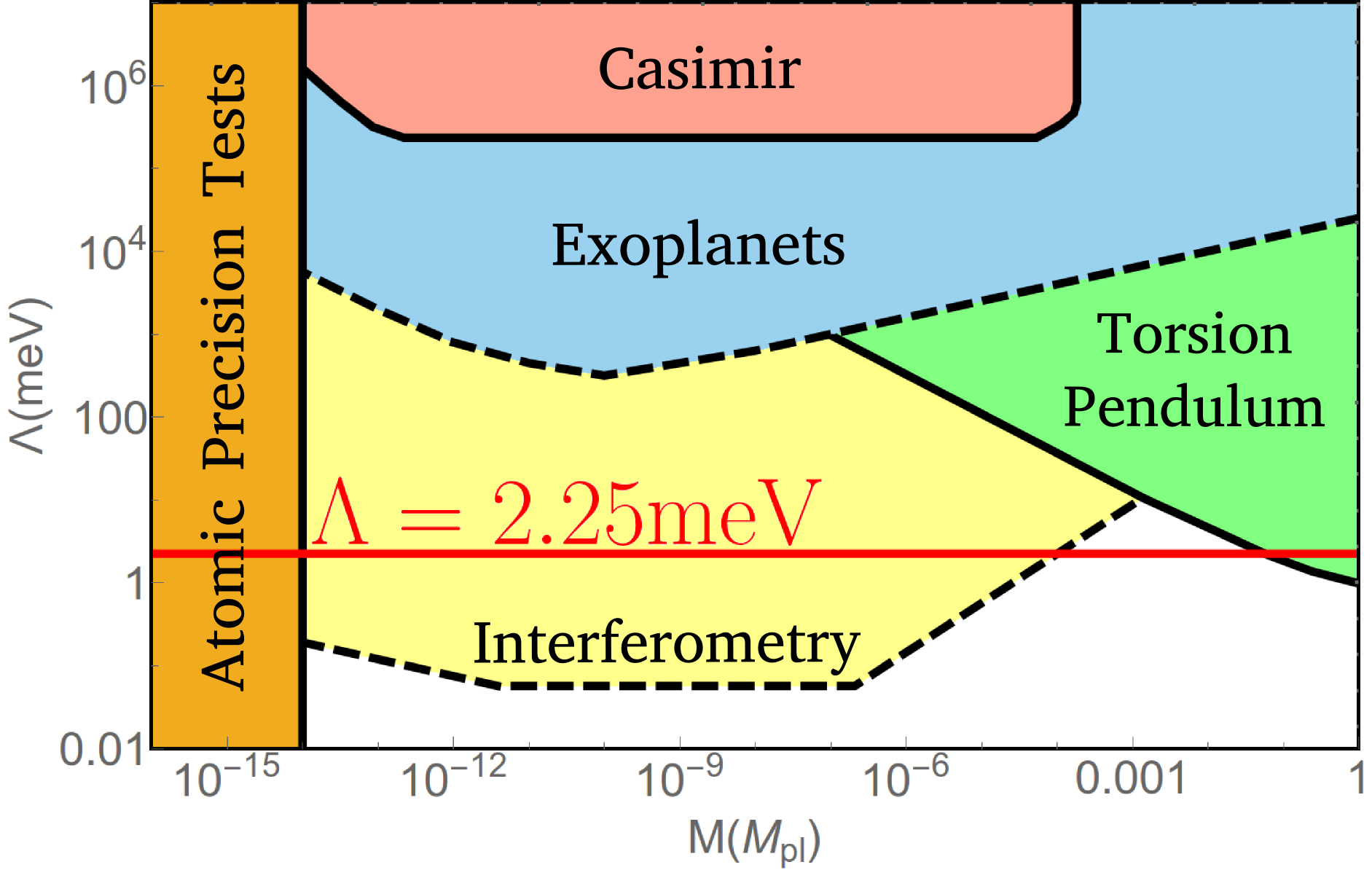}
\caption{Constraints of the Chameleon Ratra-Peebles model (blue). The excluded regions in ($M$,$\lambda$) parameter space (We fix $n=1$) compared to other experiments \citep{Brax:2007vm,Brax:2010gp,Upadhye:2012qu,Hamilton:2015zga}. The GR is recovered when $M \to \infty.$} \label{cham}
\end{figure}

\begin{figure*}[tbp]
\includegraphics[width=\columnwidth]{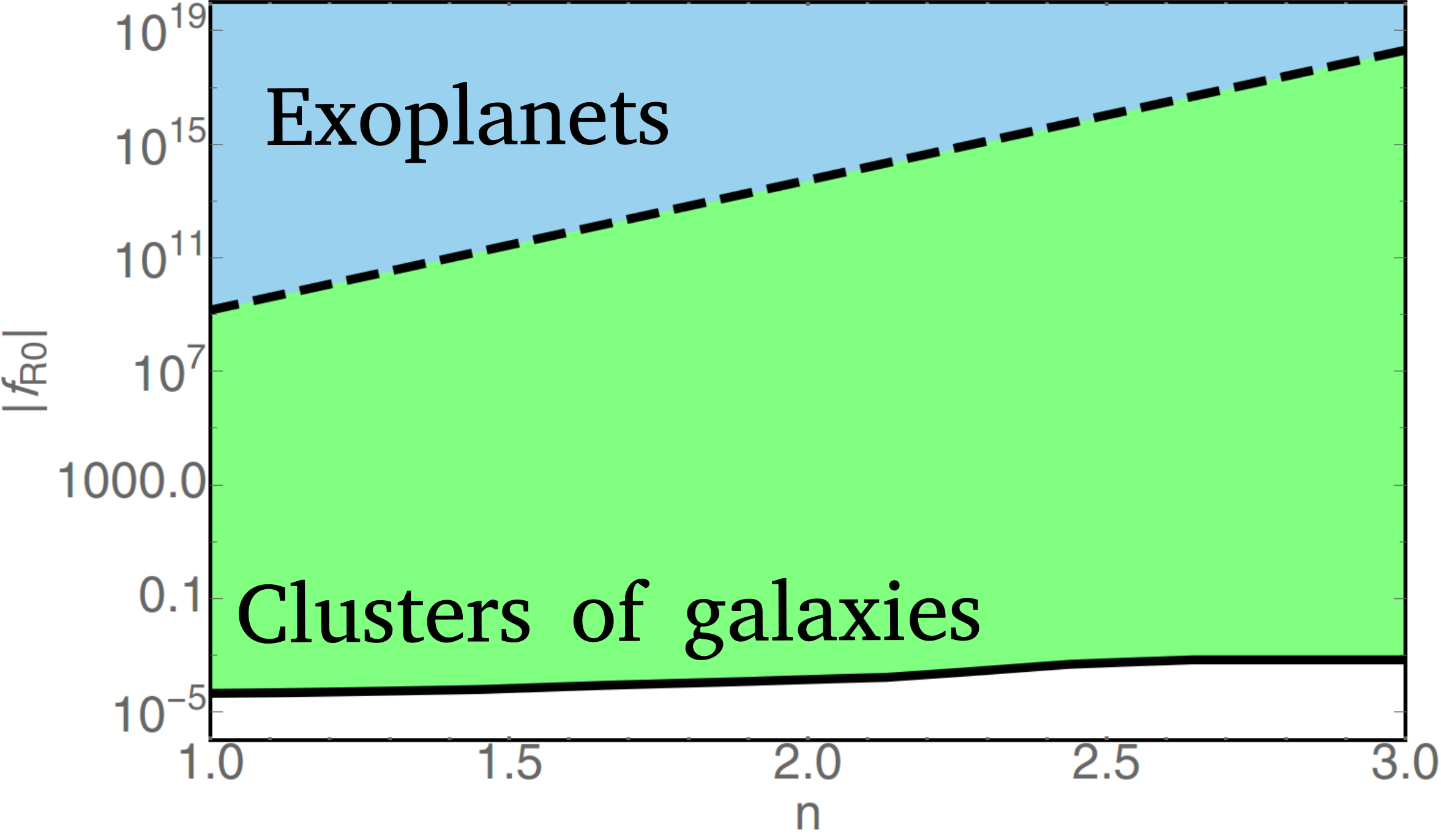}~~~~~~~~~~
\includegraphics[width=.9\columnwidth]{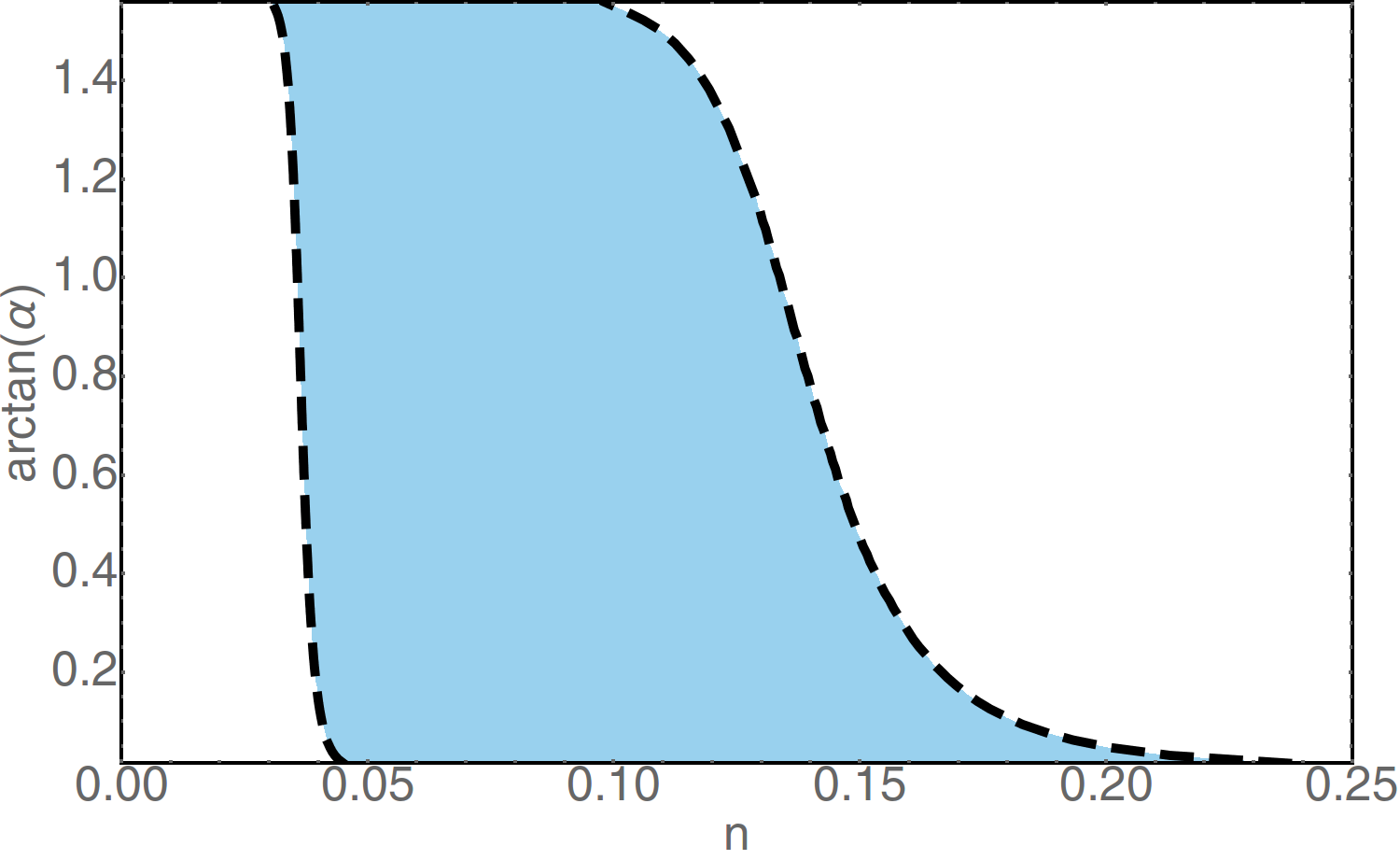}
\caption{Constraints of $f(R)$ models: (Left) Constraints for the Hu-Sawicki model from exoplanets (blue) and the results from clusters of galaxies \cite{Cataneo:2014kaa} (green). The GR is recovered when $f_{R0} \to 0$; (Right) Constraints for $\gamma$ gravity from exoplanets. Filed regions are ruled out by 95\ of confidence level. The GR is recovered when $\alpha \to 0$ or $n \to \infty$.} \label{fR}
\end{figure*}

\subsection{Yukawa potential} 
In gravitation, this potential appears when there is a massive propagator for gravitation or a scalar-field coupled to the matter fields, for example. The action of this kind of theory is
\begin{equation}
S = \int \dd^4 x \sqrt{-\mathrm{g}} \left\{\frac{M_\mathrm{pl}^2}{2}R - \frac{1}{2}(\partial\phi)^2 - V(\phi)\right\} + S_m(\tilde{g}_{\mu\nu},\Psi),
\end{equation}
where $M_\mathrm{pl} = (8\pi G)^{-1/2}$ is the reduced Planck mass, $\phi$ is the scalar field, $V(\phi)$ is the potential, $S_m$ is the matter action, $\tilde{g}_{\mu\nu}=A^2(\phi)g_{\mu\nu}$ is the Jordan frame metric, $A(\phi)$ is the conformal factor and $\Psi$ represents the matter fields. Minimizing this with respect to the field we find the Klein-Gordon equation
\begin{equation}
	\square\phi = \diff{V_{eff}}{\phi},
\end{equation}
where the effective potential is given by $V_{eff}(\phi) = V(\phi) + \rho\log A(\phi)$. Perturbing the field around the value which minimizes $V_{eff}$  ($\bar{\phi}_g$) the equation can be approximated as
\begin{equation}
	\square\phi \approx \frac{\beta(\bar{\phi}_g)}{M_\mathrm{pl}}\rho + m^2(\bar{\phi}_g) \phi,
\end{equation}
where
\begin{equation}
	\beta = M_\mathrm{pl}\diff{\log A(\phi)}{\phi},~~~~~m^2 = \frac{\dd^2 V_{eff}}{\dd \phi^2}
\end{equation}
are the coupling and the mass of the field, which is related to the force range by $\lambda=\hbar/mc$, respectively. In general, even $\beta$ and $\lambda$ are functions that depend on the parameters of the theory, and can suppress the fifth force in specific conditions. When that happens we say that the model has a screening mechanism. We begin by testing the case which $\beta$ and $\lambda$ are constants, i.e. it has no physical mechanism to suppress the fifth force. In this case we have, for a static point star of constant mass ($M_\star$) in a background of constant density ($\rho_g$), that
\begin{equation}
	(\nabla^2-m^2)\phi = \beta \frac{M_\star}{M_\mathrm{pl}} \delta^3(r).\label{kg_eq}
\end{equation}
In the non-relativistic case the test particle (planets) motion moves under the influence of a effective gravitational potential given by $\Phi = \Phi_N + \Phi_{5th}$, where $\Phi_{5th} = \log A(\phi)\simeq\beta\phi$ is the fifth force potential. From the  eq. \ref{kg_eq} solution of that is a Yukawa correction
\begin{equation}
	\Phi_{5th} = 2\beta^2e^{-r/\lambda}\Phi_N.
\end{equation}
Taking the gradient we find that the ratio between the forces is
\begin{equation}
	\varepsilon = 2\beta^2 \left(1+\frac{r}{\lambda}\right)e^{-r/\lambda}.
\end{equation}
This model has two free parameters: $\beta$, $\lambda$.

\subsection{Chameleon} This screening mechanism occurs when the field is coupled to matter in such way that the effective range of the field, $\lambda$, depends on the local matter density \citep{Khoury:2003rn,Khoury:2013yya}: when the density is low, the range is large and the fifth force acts at large scales; when density is high, the range is short, so the fifth force is screened.

As an example, we choose a field coupled to the matter via a linear conformal factor $A(\phi) = e^{\phi/M} \simeq 1 + \phi/M$. The screening happens when we consider the star as a spherically symmetric distribution of matter of constant density $\rho_\star$ embedded in medium, the galaxy, of constant density $\rho_g$. The solution outside the sphere depends both of the field inside and outside the star
\begin{equation}
	\phi(r) = \bar{\phi}_g + (\bar{\phi}_\star - \bar{\phi}_g)\frac{R}{r}e^{-(r-R)/\lambda},
\end{equation}where $\bar{\phi}_\star = \bar{\phi}(\rho_\star)$ and $\bar{\phi}_g = \bar{\phi}(\rho_g)$ are the values of the field which minimizes the effective potential $V_{eff}(\rho,\phi) = V(\phi) + \rho \phi /M$, inside the star and in the galactic medium, and the mass of the field is given by $m_\phi^2 = V_{,\phi \phi}(\bar{\phi}_g)$. Therefore
\begin{equation}
	\varepsilon = \frac{1}{M}\left( \frac{\bar{\phi}_\star - \bar{\phi}_g}{\Phi_S}\right)\left(1+\frac{r}{\lambda}\right) e^{-(r-R)/\lambda},
\end{equation}
Choosing a Ratra-Peebles potential, $V(\phi) = \Lambda^{n+4}/\phi^n$ we have
\begin{equation}
	\lambda^2 = \frac{\bar{\phi}_g^{n+2}}{n(n+1)\Lambda^{n+4}}, ~~~~~ \bar{\phi}(\rho) = \left(\frac{n M}{\rho}\Lambda^{n+4} \right)^{\frac{1}{n+1}}.
\end{equation}
This model has the following free parameters: $\Lambda$, $M$, $n$.

\subsection{f(R) gravity} This is the simplest modification that can be made to the Einstein-Hilbert Lagrangian, adding a general non linear function of $R$, $\mathcal{L} = R + f(R)$. One positive aspect of $f(R)$ is the equivalence with chameleon theories, where the scalar field is the first derivative of the $f(R)$, $\phi = f_R = df/dR$. In these models the coupling constant is $M = 2$, and the potential is 
\begin{equation}
	V(\phi) = \frac{M_\mathrm{pl}^2}{2}\frac{R f_R - f}{(1+f_R)^2}.
\end{equation}

Therefore
\begin{equation}
	\varepsilon = \left(\frac{f_{R\star}-f_{Rg}}{2\Phi_S}\right)\left(1+\frac{r}{\lambda}\right)e^{-(r-R)/\lambda},
\end{equation}
where $f_{R\star} = f_R(R_\star)$ and $f_{Rg} = f_R(R_g)$. $R_\star$ and $R_g$ are the values of the $R$ in the minimum of the effective potential inside the star and in the galactic medium respectively. In those environments, the mass of the field is high, so gravity is close to General Relativity, and we can make the approximation $R_g \approx 8 \pi G \rho_g$ and $R_\star \approx 8 \pi G \rho_\star$. From the effective potential we find that the range is $\lambda^2 = 3f_{RR}(R_g)$, where $f_{RR} = d^2f/dR^2$.

In this work we investigate two different $f(R)$ models. The first one is the model proposed by Hu and Sawicki \cite{Hu:2007nk}. For this model we have \begin{equation}
	f_R \simeq -|f_{R0}|\left( \frac{R_0}{R}\right)^{n+1},
\end{equation}in the high curvature regime, where $f_{R0}$ is the present value of $f_R$, $n$ is a positive constant and $R_0 = 3H_0^2 (4-3\Omega_{m0})$ is the effective present value of the background Ricci scalar. $H_0$ is the present value of the Hubble parameter and $\Omega_{m0}$ is the matter density parameter.

The second model is the $\gamma$ gravity model proposed in \cite{O'Dwyer:2013mza}. In this case we have \begin{equation}
	f_R = -\alpha e^{-(R/R_*)^n},
\end{equation}where $R_*$ is a positive constant, which is related to cosmological background by $R_* = 6nH_0^2(1-\Omega_{m0})/\alpha \Gamma(1/n)$.
These models have the following free parameters: $f_{R0}$, $n$ for Hu-Sawicki; $\alpha$, $n$ for $\gamma$ gravity.

\subsection{Symmetron}

\begin{figure*}[tbp]
\includegraphics[width=\columnwidth]{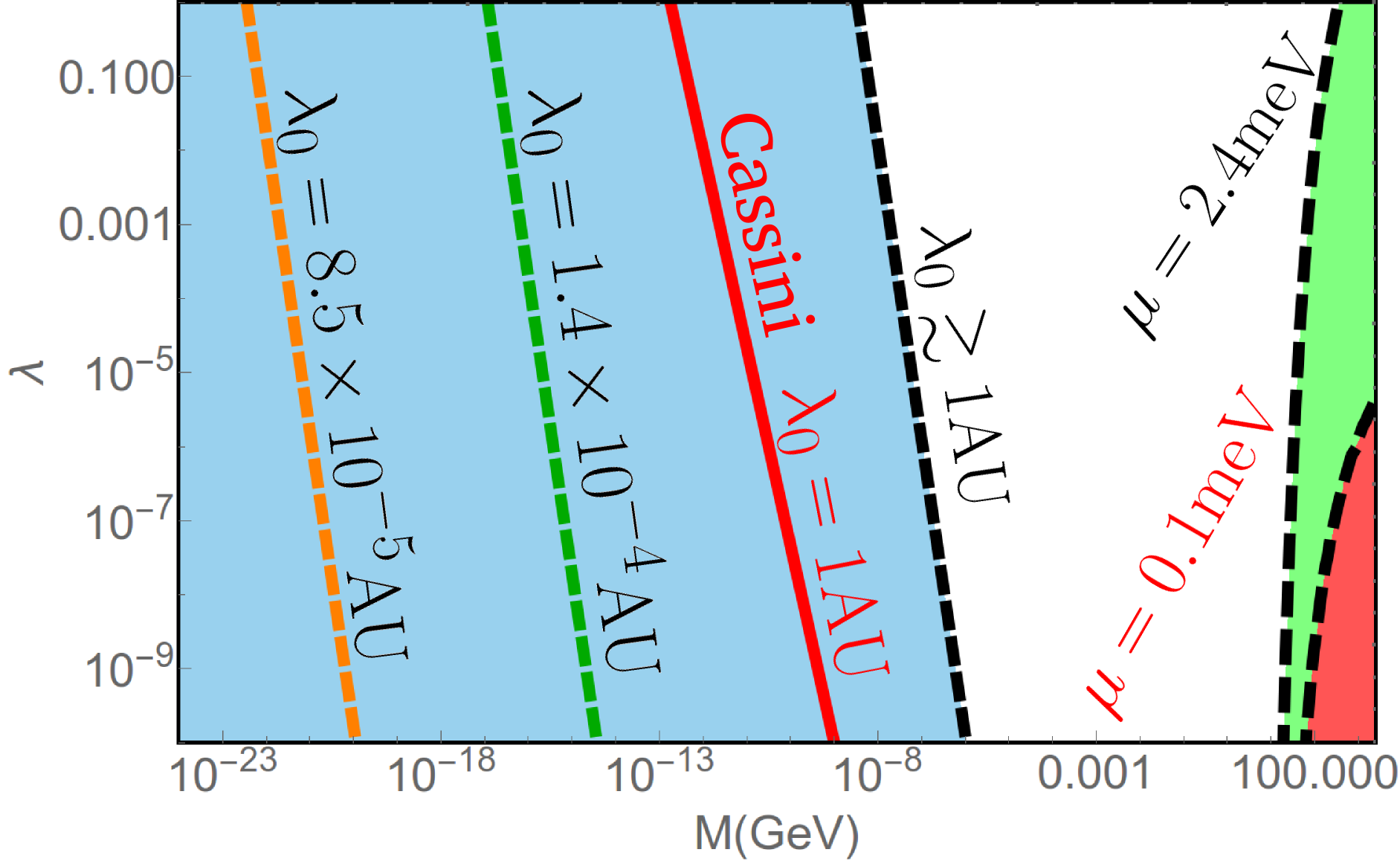}~~~~~~~~~~
\includegraphics[width=\columnwidth]{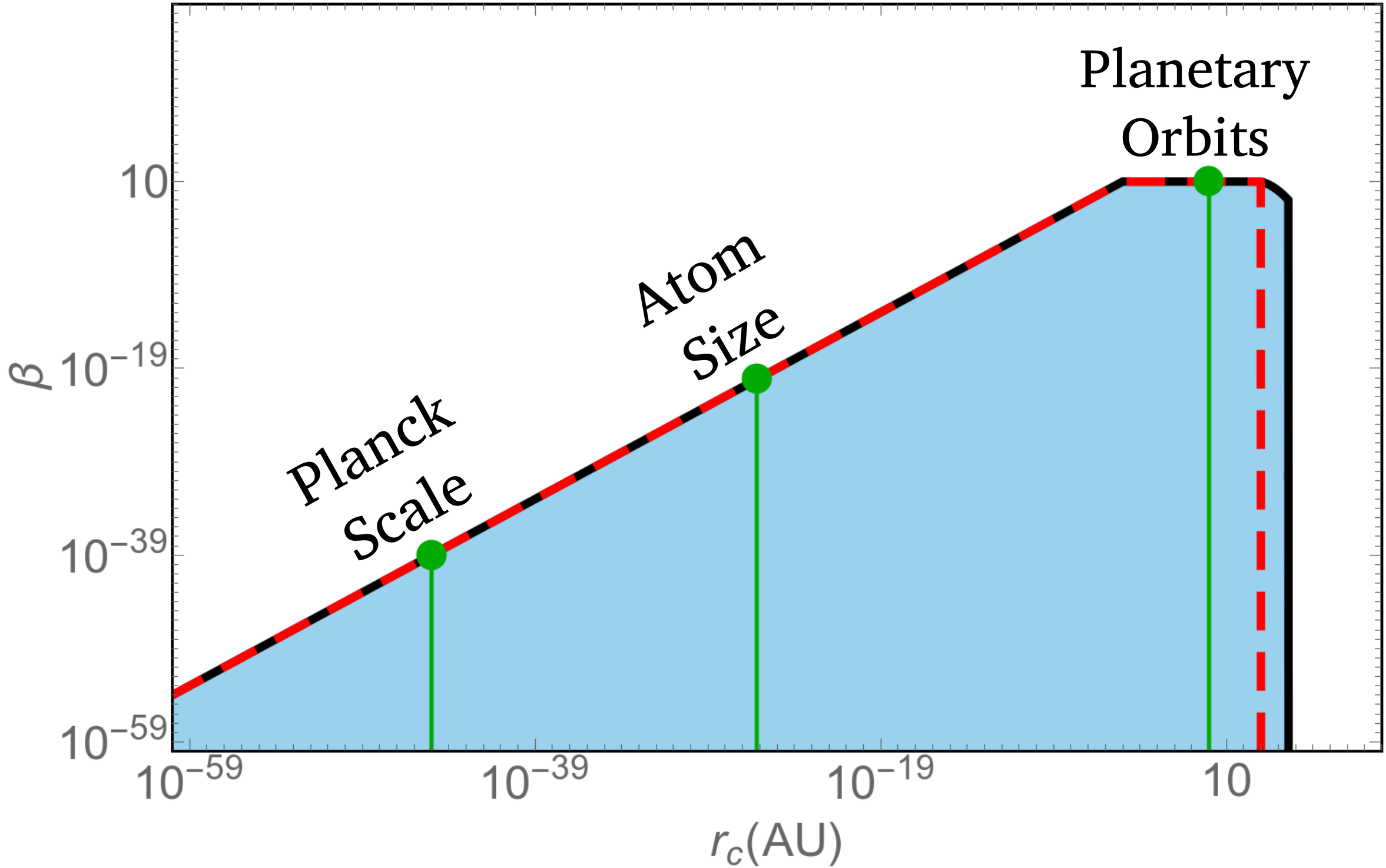}
\caption{
Left: Constraints of Symmetron; the blue region is the ruled out region for $\lambda_0\gtrsim 1\mathrm{AU}$, the green and orange dasehd lines are the bounds for smaller values of $\lambda_0$, are ruled out by 95\% \textrm{C.L.} Red line indicates the approximate bounds from the Cassini mission \citep{Brax:2012gr}. The two curves in the right side are the values ruled out by the torsion pendulum experiment in \citep{Upadhye:2012rc}. The GR is recovered when $\lambda \to \infty$ or $\mu^2 M^2 \to \rho_g$. Right: Constraints of Vainshtein screening, the blue region is ruled out by 95\% \textrm{C.L.}, the black and the red dashed lines are the bounds for the cubic only and the maximally quartic cases, respectively, note that the deviations are very small. The GR is recovered when $\beta \to 0$ or $r_c \to \infty$.}\label{symmetron_vainshtein}
\end{figure*}

The Symmetron is a matter-coupled scalar field which effectively decouples from matter in high-density regions through a symmetry restoration. This is done by introducing a potential on the symmetry breaking form \citep{Hinterbichler:2010es,Upadhye:2012rc}
\begin{equation}
	V_{eff}(\varphi) = -\frac{1}{2} \mu^2 \varphi^2 + \frac{1}{4} \lambda \varphi^4,
\end{equation}
and a quadratic conformal factor, $A(\varphi) \simeq 1 + \varphi^2/2M^2$. Which gives the follow effective potential
\begin{equation}
	V_{eff}(\varphi) = V_0 + \frac{1}{2}\left( \frac{\rho}{M^2} - \mu^2\right) \varphi^2 + \frac{1}{4} \lambda \varphi^4,
\end{equation}
From the spherical solution we get
\begin{equation}
	\varepsilon = \frac{\phi_g^2}{\Phi_S M^2}\left(1-\frac{R_\star}{r}e^{-r/\lambda_0}\right)\left(1+\frac{r}{\lambda_0}\right)e^{-r/\lambda_0},
\end{equation}
where
\begin{equation}
\lambda_0 = \displaystyle\frac{1}{\sqrt{2}\mu},~~~~~\phi_g^2 = \displaystyle\frac{1}{\lambda}\left( \frac{\rho_g}{M^2} - \mu^2\right).
\end{equation}
This model has the following free parameters: $\mu$, $M$, $\lambda$.

\subsection{Vainshtein Screening}

The Vainshtein screening works for higher dimensionally models, like Galileon or the brane world DGP model \citep{Nicolis:2008in,Dvali:2000hr}. The action for these models can be written as a brane ($\mathcal{B}$), where the RG and the matter fields are defined, embedded in a five-dimensional bulk spacetime ($\mathcal{G}$) containing a five-dimensional generalization of GR
\begin{eqnarray}
	S &=& \int_{\mathcal{B}} \dd^4 x \sqrt{-\mathrm{g}} \frac{M_\mathrm{pl}^2}{2}R + \int_{\mathcal{G}} \dd^5 x \sqrt{-\mathrm{g^{(5)}}} \frac{R^{(5)}}{16\pi G^{(5)}}+\nonumber\\
    &+& S_m\left(g_{\mu\nu},\Psi\right),
\end{eqnarray}
where the $(5)$ index denotes the quantities in the bulk spacetime. This reduces, in the weak field regime, to \cite{Andrews:2013qva}
\begin{eqnarray}
S = \int \dd^4 x \left[ \sum_{i=2}^5\frac{\alpha_i}{\Lambda^{3(i-2)}}\mathcal{L}(\phi) + \frac{\phi}{M_\mathrm{pl}}T \right] + S_{p},
\end{eqnarray}
where $\mathcal{L}_i$ is a function of $i$-th order of $\phi$ and $S_p$ is the action of the movement of point particles. Minimizing and integrating it around the star we find
\begin{equation}
	\alpha_2\left(\frac{\phi'}{r}\right) + 2\frac{\alpha_3}{\Lambda^3}\left(\frac{\phi'}{r}\right)^2 + 2\frac{\alpha_4}{\Lambda^6}\left(\frac{\phi'}{r}\right)^3 = \frac{M_\star}{4\pi M_\mathrm{pl}r^3}
\end{equation}
stability conditions imposes that
\begin{equation}
	0 \leq \frac{\alpha_2 \alpha_4}{\alpha_3^2} \leq \frac{2}{3}
\end{equation}
taking the lower limit ($\alpha_4=0$) we find the cubic only case, which the exact solution, and using that $F_{5th} = \phi'(r)/2$, we find
\begin{equation}
	\varepsilon = \frac{2}{3\beta}\left( \frac{r}{r_V} \right)^3 \left[ \sqrt{1+\left( \frac{r_V}{r} \right)^3} - 1 \right],
\end{equation}
where $\beta = \alpha_2/3M_\mathrm{pl}$ is the coupling, $r_V = \left(16 \Phi_S R_\star r_c^2/9\beta^2 \right)^{1/3}$ is the Vainshtein radius and $r_c^2 = \alpha_3/M_\mathrm{pl}\Lambda^3$ is the crossover scale. The upper limit ($\alpha_4 = 2\alpha_3^2/3\alpha_2$) gives the maximally quartic case
\begin{equation}
	\varepsilon = \frac{4}{3\beta}\left( \frac{r}{r_V} \right)^3 \left[ \sqrt[3]{1+\frac{3}{4}\left( \frac{r_V}{r} \right)^3} - 1 \right].
\end{equation}

The free parameters in these models are: $r_c$ and $\beta$.
\section{Results} The fig. \ref{yukawa} shows the ruled out region for Yukawa potential in ($\lambda$, $\alpha=2\beta^2$) parameter space. Our constraints are weaker than the ones obtained when using data from the solar system\citep{Li:2014hya}. In there the authors found an upper-bound for the coupling constant of $\alpha \leq 3.1 \times 10^{-11}$ and $\alpha \leq 5.2 \times 10^{-11}$ for $\lambda=0.15\mathrm{AU}$ and $\lambda=0.21\mathrm{AU}$ respectively. While here we find $\alpha \leq 1.5 \times 10^{-2}$ and $\alpha \leq 1.4 \times 10^{-2}$ in $95\%$ (\textrm{C.L.}) for the same values of $\lambda$ respectively. The reason for our weaker bounds is due to the fact that there are much more, and better, data of orbits around the Sun than the exoplanets data. The consequence is that their constraints are about nine orders of magnitude better than ours. This difficulty will be softened only by increasing the number of exoplanets, since the error decreases with the number of observations, $\sigma \sim 1/\sqrt{N}$. However other models, with screening, can be better constrained because the fifth force strength depends strongly on the environment physical properties.

The results for the Chameleon Ratra-Peebles model are displayed in the top panel of the figure \ref{cham}, the filled regions are ruled out in 95\% \textrm{C.L.}. The bounds from other experiments are also included \citep{Brax:2007vm,Brax:2010gp,Upadhye:2012qu,Hamilton:2015zga}, here $n$ was fixed equal to $1$ to compare with those experiments. As one can see from the figure, the exoplanets does not give us the best upperbounds for $\Lambda$, in particular the value $\Lambda = 2.25\mathrm{meV}$, which corresponds to the value of the cosmological constant predicted by the Planck satelite \cite{Ade:2015xua}, is allowed for any value of $M$. Therefore impose a background evolution close to $\Lambda$CDM is enough to ensure that the screening works in planetary scales.

Figure \ref{fR} (left) shows the results for $f(R)$ models. In bottom panel the plot the excluded region, in a 95\% of confidence level, for Hu-Sawicki model in blue. The exoplanet data provides a less restrictive constraint compared to the constraints from clusters of galaxies \citep{Cataneo:2014kaa} (also in figure \ref{fR}). The authors find  $\log_{10}|f_{R0}| \leq -4.79$ for $n=1$ in a 95.4\% confidence level, while exoplanets gives $\log_{10}|f_{R0}| \leq 2.16$. Other methods should give better constraints, such as stellar and gaseous rotation curves in dwarf galaxies \citep{Vikram:2014uza}, which gives $\log|f_{R0}| \leq -6$ or distance indicators in the nearby universe \citep{Jain:2012tn}, which gives $\log|f_{R0}| \leq -6.4$. Figure \ref{fR} (Right) shows the excluded region in parameter space for $\gamma$ gravity. These constraints are weaker than the ones from cosmological bounds, the upper-bounds for $n$ should be less than $0.25$ for any value of $\alpha$. According the \cite{O'Dwyer:2013mza} these values should not generate a final de Sitter attractor. This condition can also exclude the allowed region for small values of $n$ ($n \lesssim 0.04$). The constraints from exoplanets data allow a large region in the parameter space of these models because the information from background cosmology used already ensures that the screening works very well.

Fig. \ref{symmetron_vainshtein} (left) shows the results for Symmetron. For this models the screening is automatically satisfied, if one chooses $\mu$ and $M$ such as $\phi_g^2=0$, however we explore cases which deviate from this choices. The blue region is the ruled out region for $\lambda_0\gtrsim 1\mathrm{AU}$, the green and orange dasehd lines are the bounds for smaller values of $\lambda_0$. The red line represents the approximated bounds for the Sun surface, data from the Cassini Satelite \citep{Bertotti:2003rm}, given by the relation $A(\phi_g)-1 \lesssim 5 \times 10^{-4}\Phi_\odot$ \citep{Brax:2012gr} for $\lambda_0 = 1\mathrm{AU}$, where $\Phi_\odot \simeq 2\times10^{-6}$ is the surface gravitational potential of the Sun. Notice that our results are more restrictive, probing the power of the method especially because the field is now proportional to the square root of the deviation, i.e. $\phi_g \propto \sqrt{\varepsilon}$. We also include the bounds from a torsion pendulum experiment \citep{Upadhye:2012rc} for comparison. The torsion pendulum gives an upper-bound for $M$, while the exoplanets gives us a lower-bound.

For the Vainshtein screening mechanism, Fig. \ref{symmetron_vainshtein} (right), we allow $r_c$ and $\beta$ to be free parameters. The lower-bound for $\beta$ is larger as $r_c$ is bigger as shown in fig. \ref{symmetron_vainshtein}. For $r_c \sim 1\mathrm{AU}$ (planetary orbits), $\beta$ must be higher than $10$, in atom scale ($r_c \sim 1\mathrm{fm}$) it reduces to $1.7\times10^{-20}$ and to $2.5\times10^{-39}$ for Planck scale ($r_c\sim\ell_\mathrm{pl}$). The choice of a very large values for $\beta$ or $r_c$ ($\beta > 10$ or $r_c > 4\times 10^4\mathrm{AU}$, which includes cosmological scales $r_c \sim H_0^{-1}$) is enough to make the screening works in planetary scales, suppressing any modifications to General Relativity. The difference between the cubic only and the maximally quartic cases are very small.

\section{Conclusions and perspectives} In this Letter we show a promising, novel and complementary method to constrain modified gravity using extrasolar planets data. Using the orbits of exoplanets we are able to put tighter constraints on the Chameleon models and the Symmetron. These bounds are comparable and competitive with the ones from other experiments, because exoplanets systems have a wide range of masses for the host star, resulting in different screening scales and effectiveness. For Vainshtein screening we find a relation between the free parameters ($\beta$, $r_c$) beyond the trivial choice $r_c \sim H_0^{-1}$.

\section*{Acknowledgements}

MVS acknowledges the Brazilian agencies FAPERJ, CAPES and CNPq; and Sophie Harris-Edmond for useful suggestions. DFM acknowledges the Research Council of Norway.

This research has made use of the Exoplanet Orbit Database and the Exoplanet Data Explorer at \url{exoplanets.org}.

\bibliographystyle{model1-num-names}
\bibliography{bibfile.bib}

\end{document}